\pdfoutput=1
\documentclass[manuscript]{aastex6}		% For AJP manuscript journal format
\usepackage{epsfig}			% For eps figures, old commands
\usepackage{graphicx,color}		% For eps figures, newer & more powerfull
\usepackage{amssymb}			% For useful mathematical symbols
\usepackage{color}			% For color text: \color command
\usepackage{amsmath}			% Provides various features to facilitate writing math formulas
\usepackage{rotating}			% For rotate any object of an arbitrary angle
\usepackage{float}			% For improving the interface for defining floating objects
\usepackage{textcomp}			% Pro­vides many text sym­bols (such as baht, bul­let, copy­right,etc )
\usepackage{psfig}
\usepackage{times}
\begin{document}

\title{Butterfly Diagram and Carrington Maps for Century-Long Ca~{\sc ii}~K Spectroheliograms from Kodaikanal Observatory}
\author{Subhamoy Chatterjee$^{1}$,
Dipankar Banerjee$^{1, 2}$,
B. Ravindra$^{1}$
}
 
\affil{$^{1}$Indian Institute of Astrophysics, Koramangala, Bangalore 560034, India. e-mail: {\color{blue}{dipu@iiap.res.in}}\\
$^{2}$Center of Excellence in Space Sciences India, IISER Kolkata, Mohanpur 741246, West Bengal, India  \\}
%\date{}							% Activate to display a given date or no date

%\maketitle

\begin{abstract}
The century-long (1907-2007) Ca~{\sc ii}~K spectroheliograms from Kodaikanal Solar Observatory (KSO) are  calibrated, processed and analysed in the present study to follow the evolution of bright on disc structures called plages, the possible representatives of  magnetic activity on the Sun. This has been the longest dataset studied in Ca~{\sc ii}~K till date covering about 9.5 cycles of 11 year periods. Plages were segmented with area $\geq 1\:\textrm {arcmin}^2$ using global thresholds for individual full disc images and subsequent application of morphological closing operation.  Plage index was calculated and seen to have close positive correlation with fractional disc area covered by plages. The newly generated plage area cycle (from KSO) was compared with the same from Mount Wilson observatory (Correlation~$95.6\%$) for the overlapping years i.e. 1915-2000. Study illustrated time-latitude distribution of plage centroids rendering butterfly diagram (as observed for sunspots). The 3D visualisation of the diagram showed one to one mapping between plage location, time and area. This work further delineated positional correlation between  magnetic patches and plage regions  through comparison of synoptic maps derived from both Kodaikanal Ca~{\sc ii}~K images and space based full disc LOS (line of sight)  magnetograms.  Regular synoptic magnetograms from ground based observatories are available only after 1970s. Thus the long term Ca~{\sc ii}~K data from KSO can be used as a proxy for estimating magnetic activity locations and their strengths at earlier times.
\end{abstract}


\section{Introduction}
Astronomical objects like stars are embedded with self organising system behaviour which is  expressed in different scales and modes over time. Multimodal long duration data may assist in gaining insight about such complex system dynamics. Being a part of the solar system it is of particular interest for us to study the evolution of spatially resolved features on the Sun. %Thus, sun, being a reference young star, provides options to evaluate its features at different radial heights in the context of its evolution. %
 In this regard, long term data available from different ground based observatories provide valuable informations at different wavelengths. Ca~{\sc ii}~K (3933.67 \AA),  line of singly ionised  calcium, is one such study wavelength. This absorption line extends facts about solar chromosphere \citep{stix2004sun}. Intensities observed through this line are dependent on magnetic field strength. Locations of higher magnetic fields are correlated with bright regions in Ca~{\sc ii}~K solar intensity images \citep{2011ApJ...730...51S,1998ASPC..140..155H}. Thus the Ca~{\sc ii}~K line emission provides a good proxy for line of sight magnetic field fluxes, particularly for a historical period where regular direct magnetic field measurements are not available \citep{1975ApJ...200..747S,1989ApJ...337..964S}.
The Kodaikanal Solar Observatory (KSO) has archived  full disc Ca~{\sc ii}~K  spectroheliograms for more than a century (1906 to mid 2007), using photographic plates  as recorded  through a telescope having a 30 cm objective with f-ratio of f/21 \citep{2014SoPh..289..137P}. The effective spatial resolution was about 2 arcsec for majority of the documentation time. 
Recently, 16 bit digitisation has been performed on those plates through a CCD sensor (pixel size 15 micron cooled at $-100^o$C) to generate $4096\times4096$ raw images. This data provides a significant temporal window \citep{Foukal2009} to study the evolution of solar activity while studying the variation of  disc structures, namely plages, filaments, sunspots \citep{zhar}. While comparing the quality of different Ca~{\sc ii}~K images as recorded from Arcetri, Kodaikanal, and Mount Wilson observatories \citet{0004-637X-698-2-1000} concluded that KSO provides a homogeneous and the longest series. Note that this comparison was made based on earlier low resolution scanned images (as obtained from KSO) and  the need for higher resolution digitisation was pointed out by \citet{0004-637X-698-2-1000}. First results from the new digitisation process was reported in \citet{2014SoPh..289..137P} using KSO data on Ca~{\sc ii}~K for the period 1955-1985. We have  recalibrated  the data for the full 100 years and are presenting the first results from the new time series. The extracted data through this study is also being made available to the scientific community for further processing. The digitised data is available through \url{https://kso.iiap.res.in/data}.

Few earlier studies have reported identification of chromospheric features, to generate time series for the understanding the long term variation of their sizes and location. But none of them used century-long data, which allows us to study several solar cycles and inter-cycle variation and this may help in our understanding of the  the solar dynamo model \citep{0004-637X-806-2-174}. In the context of data processing, detection methods typically rely on intensity contrast and size of target features. However, contrast may differ between images causing mis-detection of structures. Also, such techniques suffer from fragmentation of segmented features resulting in centroid detection error \citep{zhar}. %An earlier work has been performed by \citet{2014SoPh..289..137P} using KSO data on Ca~{\sc ii}~K for the period 1955-1985.
In the present work we used newly digitised higher resolution century-long KSO Ca~{\sc ii}~K data for automated plage detection. This generated a number of time series (viz. area cycle, butterfly diagram and their combination) to facilitate  the study of solar activity variation keeping in mind the limitations and constraints. Section 2  illustrates the  image statistics, calibration, enhancement and processing techniques. The data was also exploited to generate synoptic Carrington Rotation (CR)  maps. Section 3 depicts plage area cycle, butterfly diagrams and positional correlation of some Ca~{\sc ii}~K CR maps with the same generated from line of sight magnetograms.  The CR maps corresponding to the 100 years data are available  online (supplementary material). We discuss the relevance of our results and present our conclusions in Section 4.

\section{Methods}
\begin{figure}[h]
\centering
  \includegraphics[width=1\linewidth]{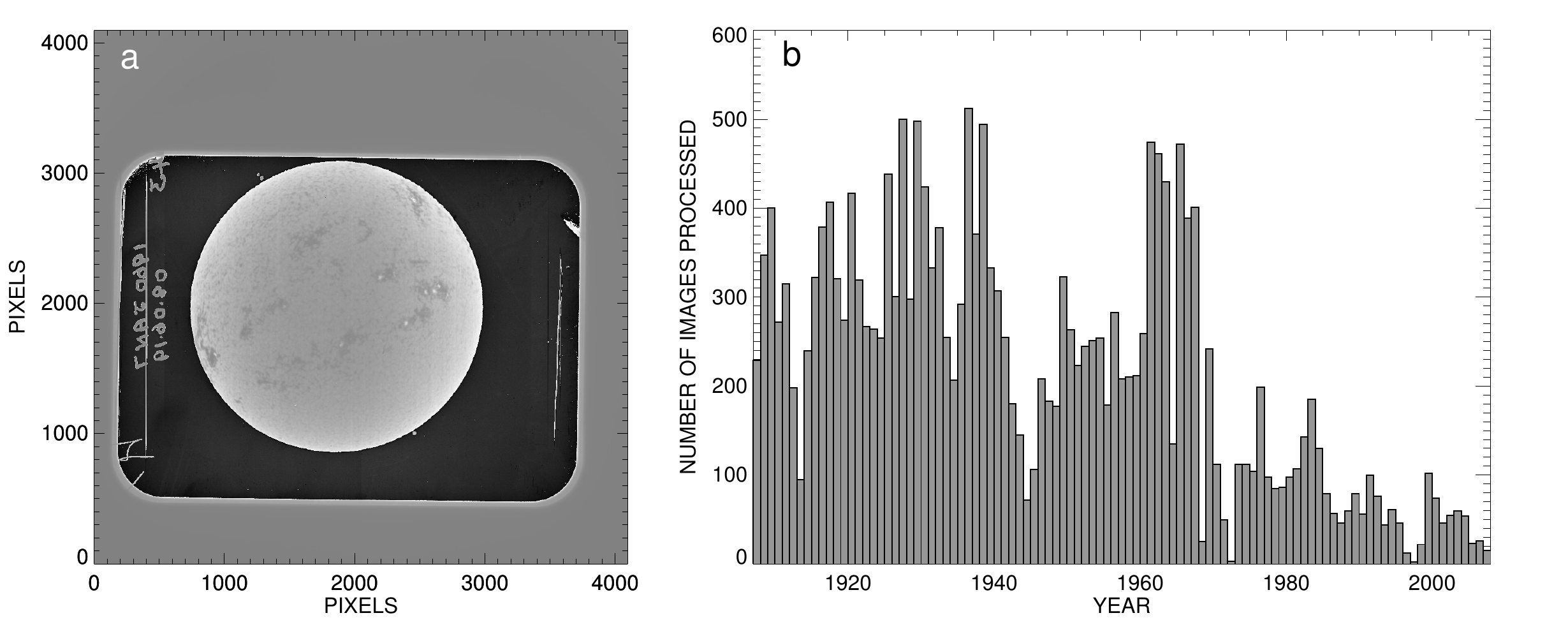}
  \caption{Input for the processing. a) One representative Ca~{\sc ii}~K spectroheliogram raw image from KSO; b) Ca~{\sc ii}~K image histogram showing number of images considered for processing year wise}
  \label{fig:CaK_stat}
\end{figure}

The digitised full disc century-long (1907 to mid 2007) 4096$\times$4096 Ca~{\sc ii}~K solar images ($\approx 0.8036$ arcsec/pixel)  from KSO (Figures~\ref{fig:CaK_stat}a-b) used in the study were recorded through a telescope having 30 cm objective with f-ratio of f/21 \citep{2014SoPh..289..137P} and spatial resolution about 2 arcsec. We employed few calibration steps before the feature detection was performed. They were namely flat fielding, disc detection and centring, north-south rotation correction, intensity inversion and limb darkening correction. Hough circle transform \citep{sonka2014image} was applied on edge detected raw images to efficiently identify the disc centre and radius in contrast to manual method described in earlier work by \citet{2014SoPh..289..137P}. The radius varied by some number of pixels systematically over time due to change in sun-earth distance during each revolution. All the disc centred images were rescaled to one fixed size ($N\times N$) having uniform disc radius ($R$) (Figure~\ref{fig:CaK_processing}a). 
\subsection{Limb darkening correction}
Limb darkening is a systematic large scale  variation of solar intensity from its centre to limb due to line of sight effect. This should ideally be a radially symmetric function but due to atmospheric and instrumental effects it loses the symmetry. In contrast to the radially symmetric polynomial method of limb darkening estimation for KSO data as presented in \citet{2014SoPh..289..137P}, current study captured the variation applying a 40 $\times$ 40 2D median filter \citep{zhar,2010SoPh..264...31B} on the resized 512 $\times$ 512 disc centred images. This essentially performed blurring to represent large scale intensity variation (Figures~\ref{fig:CaK_processing}b-c). Subsequently, the filtered image was again resized to original  4096 $\times$ 4096 size and was used to correct for limb darkening in the original image (Figure~\ref{fig:CaK_processing}d). 
\subsection{Plage detection}
\begin{figure}
\centering
  \includegraphics[width=0.8\linewidth]{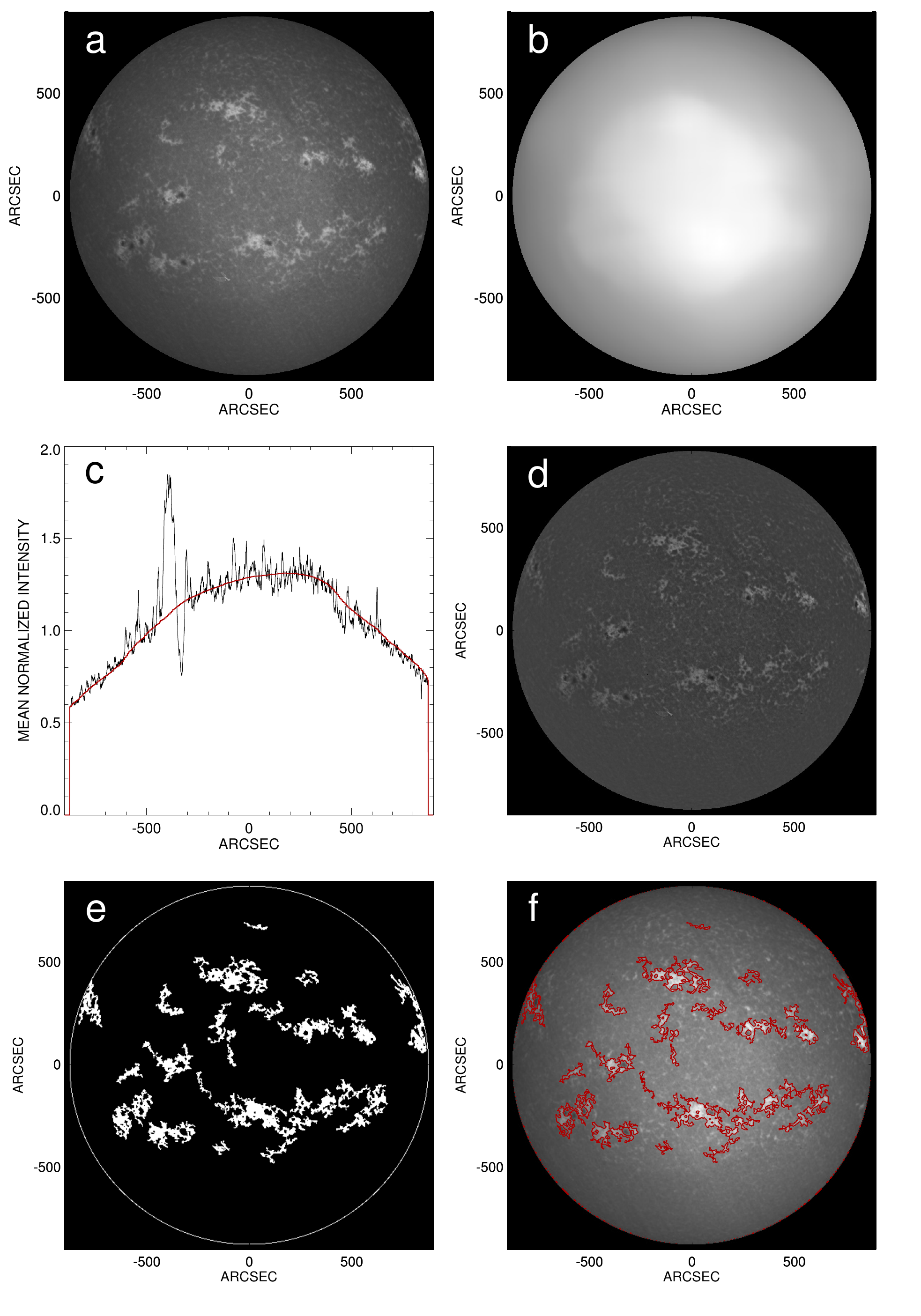}
 % \vspace{-.04\textwidth}
  \caption{Processing steps involved for a representative KSO Ca~{\sc ii}~K image to detect plages. a) Ca~{\sc ii}~K disc centred image; b) Blurred image applying large window median filtering on (a) to understand centre to limb intensity variation; c) x-profile overplotted through centre of (a) and (b); d) Limb darkening corrected image; e) Plage segmented image through row wise thresholding; f) Plage contour overplotted on (a)}
 \label{fig:CaK_processing}
\end{figure}

Histogram equalisation was applied on the limb darkening corrected images ($LDC$) to bring uniformity in intensity contrast throughout the study time and to enhance the target features with respect to background. There was no absolute intensity calibration or full disc flux measurement \citep{1988ApJ...328..347F} unlike \citet{0004-637X-698-2-1000}. The objective was only to detect plages. Global thresholds were then applied on the histogram equalised images to generate binary images ($BW$) with segmented regions of interest.  Enhanced limb darkening corrected images were characterised by disc region intensity median ($\textrm{median}_{\textrm{Disc}}$) and standard deviation ($\sigma_{\textrm{Disc}}$). The disc region was described by pixels ($i,j$) satisfying  $\sqrt{(i-\frac{N}{2})^2+(j-\frac{N}{2})^2}\leq R$. The segmentation scheme was as follows-
\begin{align}
      BW_{ij} & =1~~\:\textrm{when}\:{LDC^{\textrm{HistEqual}}}_{ij}> \textrm{median}_{\textrm{Disc}}+\sigma_{\textrm{Disc}}\nonumber\\
                   & =0~~\:\textrm{otherwise}\nonumber
\end{align}
After segmentation an area threshold of $0.25\:\textrm{arcmin}^2$ pixels was applied to discard unwanted regions. Subsequently, morph closing (a mathematical morphology operation consisting of dilation followed by erosion) \citep{sonka2014image} was applied with a $3\times3$ (cross) structuring function to get rid of fragmentation in structures of interest (Figures~\ref{fig:CaK_processing}e-f).   

\subsection{Plage index calculation}
\begin{figure}[h]
\centering
  \includegraphics[width=1\linewidth]{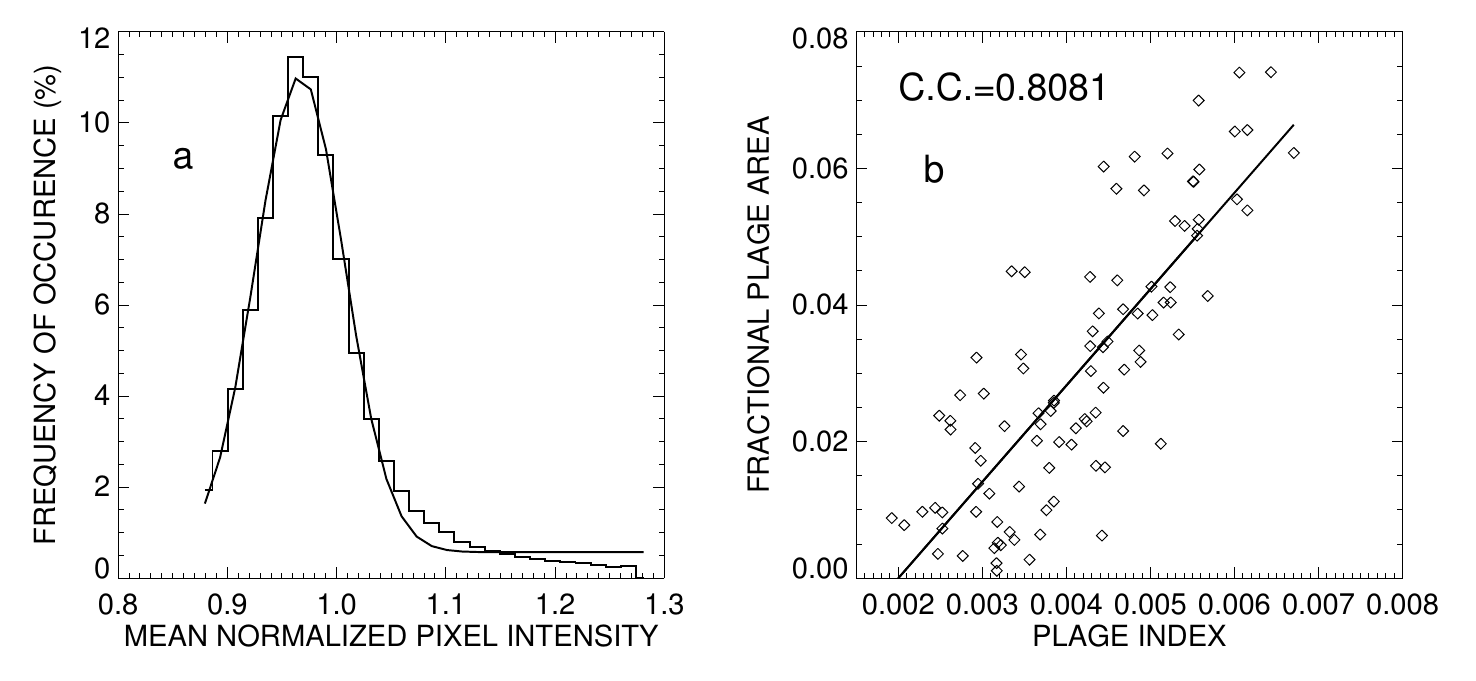}
  \caption{Validation of the current detection strategy. a) Intensity histogram of a Ca~{\sc ii}~K image disc region and its Gaussian fit; b) Correlation of fractional plage area and plage index with least square linear fit is given by, $\:\:\textrm{Fractional Plage Area}=-2.82\times10^{-2}+14.1133\times\textrm{Plage Index}$}
  \label{fig:CaK_index}
\end{figure}

\begin{figure}
\centering
  \includegraphics[width=1\linewidth]{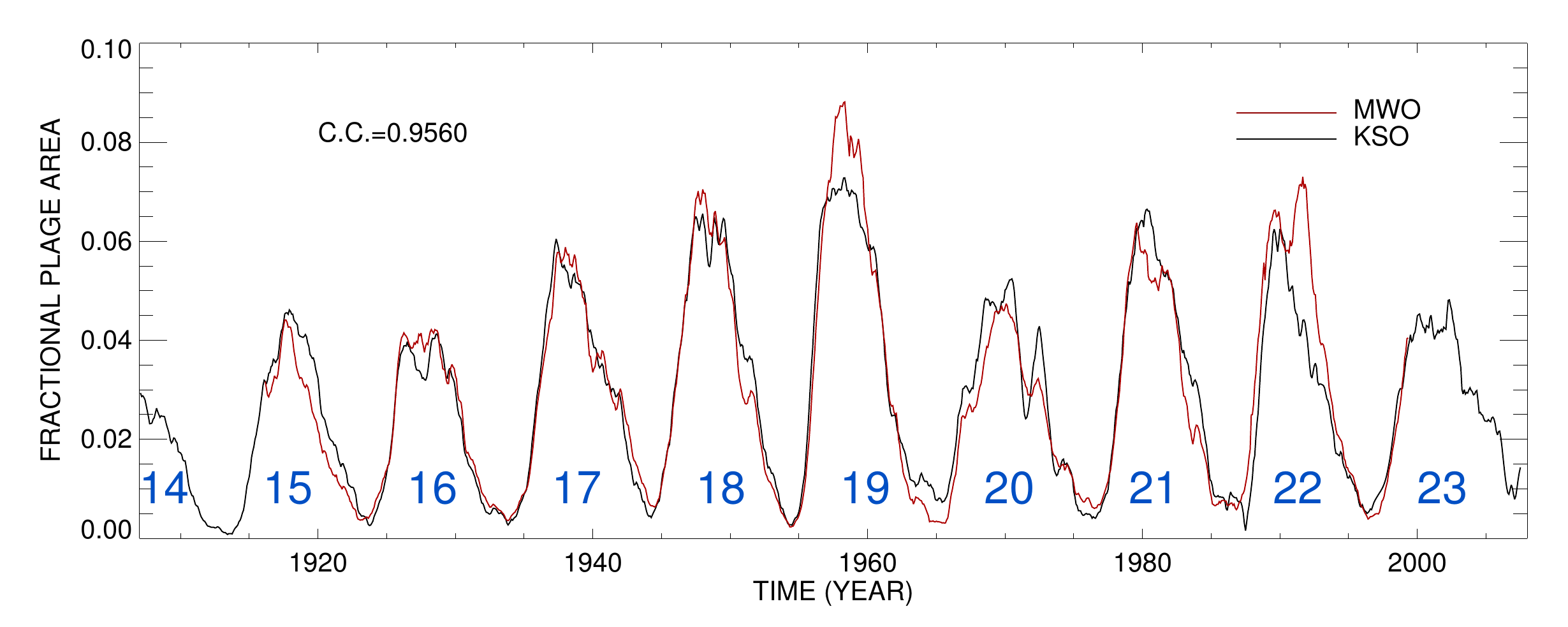}
 % \vspace{-.06\textwidth}
    \caption{Ca~{\sc ii}~K plage area cycle depicting 12 months running averaging applied over monthly average plage area. Black curve corresponds to  the Kodaikanal Ca~{\sc ii}~K archival data and the red curve corresponds Mount Wilson Observatory. Least square linear fit for the overlapping region (1915-2000) is give by, $\:\:\textrm{Plage Fraction}_\textrm{MWO}=-0.0009+1.0252\times\textrm{Plage Fraction}_\textrm{KSO}$. Cycle numbers are marked in blue.} 
  \label{fig:area}
\end{figure}

Plage index is an established metric, reflecting the occurrence of plage regions in Ca~{\sc ii}~K image. Following the description about plage index calculation as in \citet{2010SoPh..264...31B}, from the mean normalised intensity histrogram of solar disc with a bin-width of 0.01, 35 bins were selected above and below the mode. Within this range a Gaussian was fitted to calculate mean ($\mu$) and variance ($\sigma$).  
Within $\mu-2\sigma$ and  $\mu+7\sigma$, 30 bins were selected and another Gaussian ($f(x)=ke^{-\frac{(x-A)^2}{B}}+C$ where $(\mu-2\sigma)\leq x \leq (\mu+7\sigma)$) was fitted to it (Figure~\ref{fig:CaK_index}a). The constant term ($C$) of the Gaussian is named as plage index indicating the asymmetry in the histogram wings due to occurrence of high intensity regions called plages. For further validation  the correlation of plage index  with fractional plage area acquired from plage detection is also presented (Figure~\ref{fig:CaK_index}b). Hereafter for solar cycle variation studies we will use the fractional plage area, the fraction of solar disc area covered by plages, as a most favoured proxy. The 12 month running average smoothing was performed on monthly averaged plage area cycle and compared  with same for Mount Wilson Observatory (MWO) in Figure~\ref{fig:area}.

%\vspace{-.08\textwidth}
\subsection{Plage centroid detection and butterfly diagram}
After morph closing operation on binary plage detected images the individual connected structures were identified through giving certain labels to the pixels falling in same region. Subsequently, centroid latitudes for all those regions were calculated for each image and were stacked over time to generate butterfly diagram  with four area thresholds for the plage regions  $1\:\textrm{arcmin}^2$, $4\:\textrm{arcmin}^2$, $7\:\textrm{arcmin}^2$ and $10\:\textrm{arcmin}^2$ (Figure~\ref{fig:butterfly}).

%\vspace{-.03\textwidth}
  \subsection{Carrington map generation}
\begin{figure}
\centering
  \includegraphics[width=0.75\linewidth]{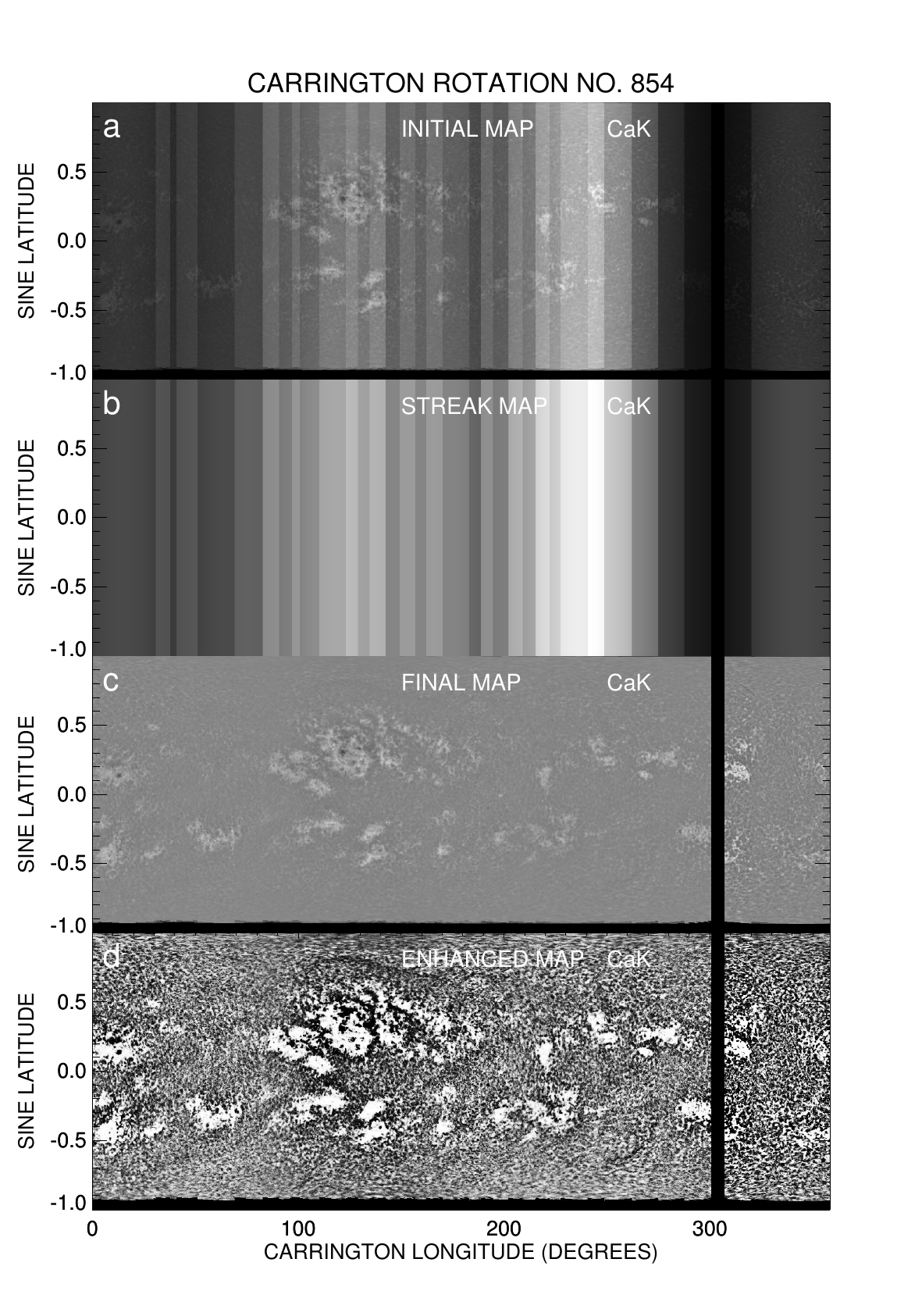}
    \caption{Carrington maps generated from Kodaikanal observatory Ca~{\sc ii}~K full disc spectroheliograms starting at 23rd July,1917.  a) Carrington map before overlap correction; b) Streak map to determine overlaps; c) Corrected map by division of (a) by (b); d) Intensity enhanced map through histogram equalization.}
  \label{fig:cr_CaK_streak}
\end{figure}

Carrington map is the Mercator projection of the spherical sun mostly generated from daily observation for one solar rotation. In this study, 60 longitude bands across the limb darkening corrected full disc Ca~{\sc ii}~K  images were selected, $B_0$ angle corrected and stretched in form of a rectangle with a weightage of fourth power of cosine \citep{2011ApJ...730...51S,1998ASPC..140..155H} over each longitude with respect to the central meridian. These slices were shifted and added according to date and time for 27.2753 days to generate a full 360 degree map of sun. A similar 360 degree map was obtained from rectangular binary slices called streak map \citep{2011ApJ...730...51S}. The overlap of same Carrington longitudes was removed through division of the original 360 degree map with streak map to form an image called Carrington map or Synoptic map (Figures~\ref{fig:cr_CaK_streak}a-d). Carrington maps were also generated from few MDI LOS full disc magnetograms available after 1996. Binary Carrington maps were generated for Ca~{\sc ii}~K and MDI using $\textrm{mean} +\sigma$ and $\textrm{mean} \pm\sigma$ thresholds respectively. These maps were further blurred using Gaussian filter to account for the feature boundary uncertainty and were cross-correlated for different relative x-y shifts. 
\vspace{-.09\textwidth}
\vspace{.1\textwidth}
 \section{Results}
For the study of the century-long magnetic activity of the Sun, the Ca~{\sc ii}~K plage structures (Figure \ref{fig:CaK_processing}), with projected area $\geq 1\:\textrm{arcmin}^2$, have been used as the primary building block for magnetic proxy. Their segmentation endorsed the area threshold as $0.25\:\textrm{arcmin}^2$ to minimise loss of fragmented plage structures before morphological closing operation. This enabled us to deal with a good number of plage structures and to achieve desired statistics for the study of solar cycle variation. Histogram equalisation on individual limb darkening corrected solar images  resulted in uniformity in image contrast. This also allowed us to detect the features automatically without any human bias of selection by visual inspection.
\begin{figure}
\centering
  \includegraphics[width=1\linewidth]{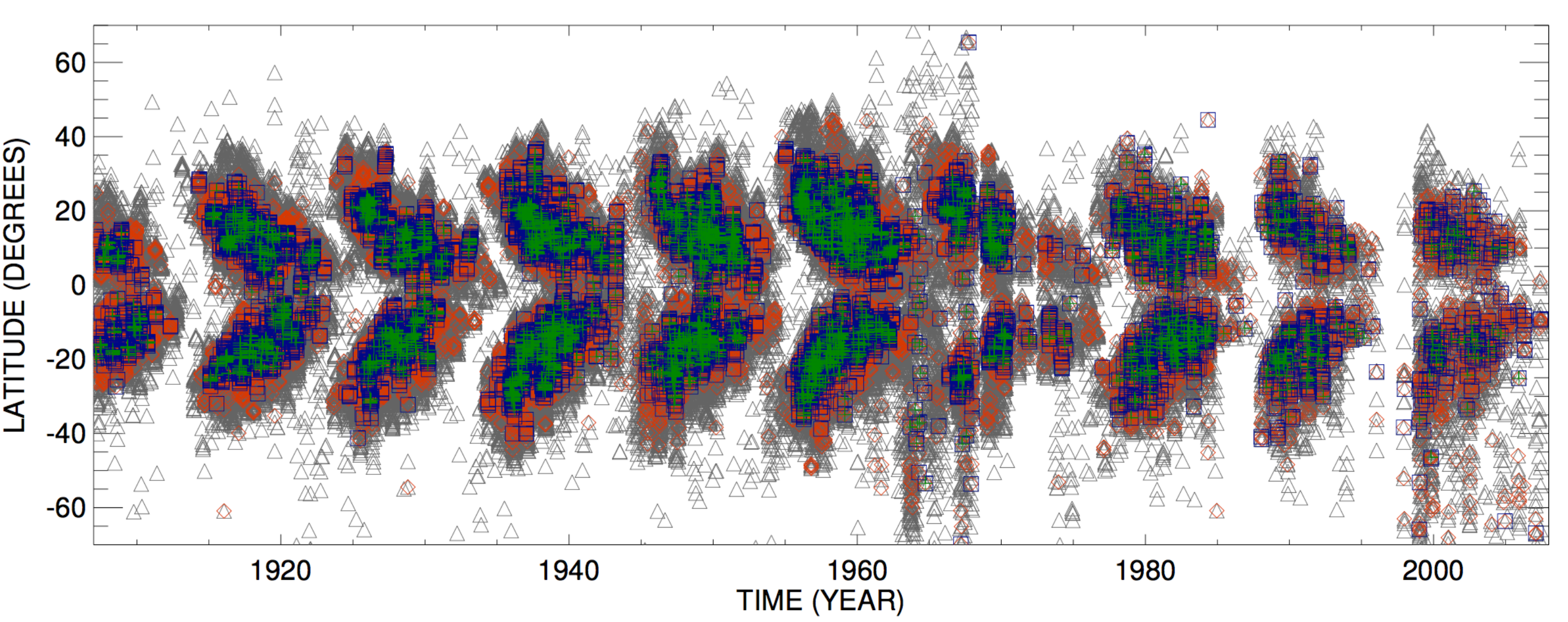}
  \vspace{-.06\textwidth}
    \caption{Ca~{\sc ii}~K butterfly diagram. Grey triangles depict centroids of plages having area $\geq 1\:\textrm{arcmin}^2$. Red, blue and green symbols depict centroids of plages having area $\geq 4\:\textrm{arcmin}^2$,  $\geq 7\:\textrm{arcmin}^2$ and  $\geq 10\:\textrm{arcmin}^2$ respectively.}
  \label{fig:butterfly}
\end{figure}
%---------
% \vspace{-.1\textwidth}
  \begin{figure}[ht]
%\hspace{-.2\textwidth}
  \includegraphics[width=1\linewidth]{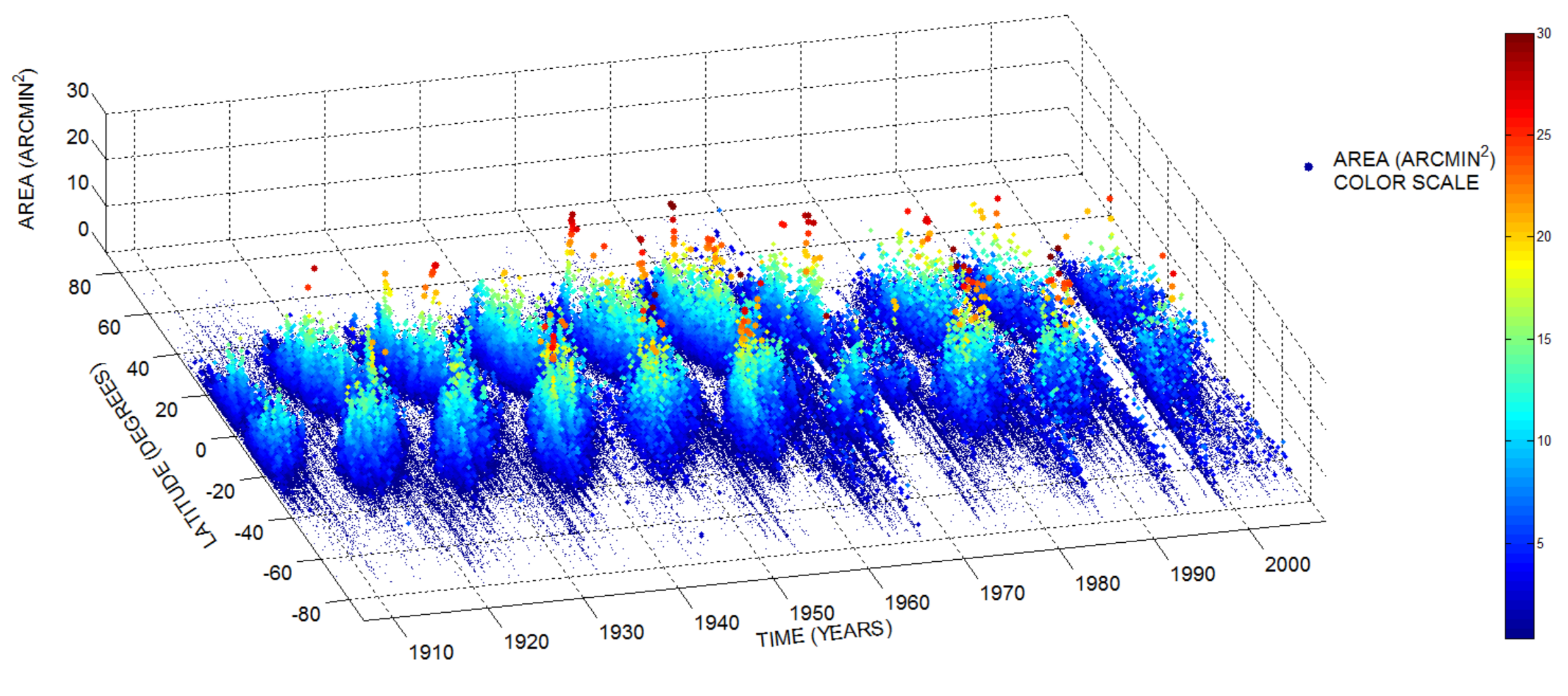}
   \vspace{-.06\textwidth}
       \caption{3 dimensional visualisation of butterfly diagram with area of individual plages as the z-axis. Minimum to maximum plage area range is defined by dark blue to dark red through green, yellow and orange as indicated by the color scale. (A dynamic version of this figure in the form of an animated movie is available online). }
  \label{fig:butterfly_3d}
\end{figure}
The area cycle, a temporal evolution of fractional disc area covered by plages, showed a good correlation ($95.6\%$) with the same from MWO for the overlapping time period (Figure~\ref{fig:area}). This is an improvement over the earlier report by \citet{2014SoPh..289..137P} on KSO Ca~{\sc ii}~K data for the period 1955-1985 in which the correlation was shown to be 89$\%$. It further revealed the cycle 19 as the strongest one (Figure~\ref{fig:area}) and evidenced the double peak behaviour in some of the cycles. Some unreal spikes were also observed at the cycle minima after 1975 due to less availability and inferior quality of images being correlated with the occurrence of outlier points in the butterfly diagram for some study period after 1985 ($\approx$2 years) and 1994 ($\approx$1 year). Those data points were removed using a criteria involving the maximum latitude of plage appearance at a certain phase of a cycle.
	
To explore the change in latitudinal distribution of enhanced Ca~{\sc ii}~K emission with time we have produced the butterfly diagram. The centroids of plages have been used for the generation of this diagram,  in contrast to \citet{2011ApJ...730...51S} who used the Carrington maps for the same. Butterfly diagram with three area thresholds (Figure~\ref{fig:butterfly}) reveals that higher plage areas are more concentrated to the equator with a spatial correlation with active regions \citep{stix2004sun}. In order to have a better visualisation of the latitudinal variation of different sizes with time a 3D version of the butterfly diagram was created, adding individual projected plage area as the third dimension (Figure~\ref{fig:butterfly_3d}).  An animated figure has also been created to demonstrate how the time-latitude distribution changes with variation of plage size. This dynamic illustration has been provided as a supplementary material in the link \url{ftp://ftp.iiap.res.in/subhamoy/cak_carrington_maps_kodaikanal/butterfly_cak_3d.gif} and is available online. %Also to be noted from Figure~\ref{fig:butterfly_3d} is that the larger flux concentrations (larger plage areas) are always closer to equator.
 Figures~\ref{fig:butterfly} \& \ref{fig:butterfly_3d} can be readily compared with  well known butterfly diagrams as generated from sunspot areas showing higher latitude to equatorward occurrence of plage structures for each cycle. It should be noted that  because of area thresholding the smallest features like network bright points are not detected in the current analysis and the poleward migration is not observed \citep{2014ApJ...793L...4P}. 
 
\begin{figure}
%\vspace{-.2\textwidth}
\centering
  \includegraphics[width=1.\linewidth]{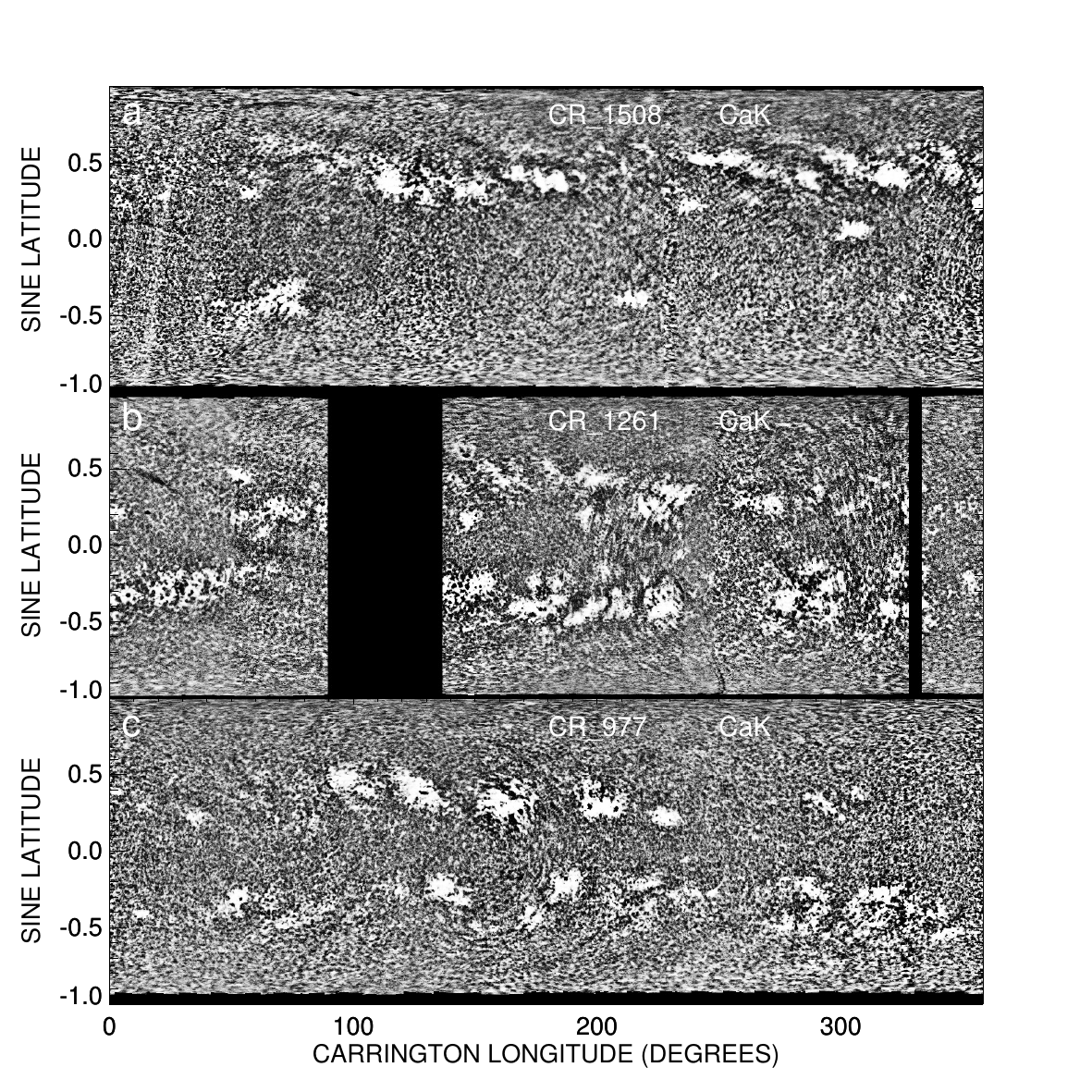}
 % \vspace{-.03\textwidth}
    \caption{Carrington maps generated from Kodaikanal observatory Ca~{\sc ii}~K full disc spectroheliograms.
   a) Carrington map starting at 25th May, 1966; b) Carrington map starting at 14th December, 1947; c) Carrington map starting at 29th September, 1926.}
  \label{fig:cr_CaK}
\end{figure}

\begin{figure}
%\vspace{-.2\textwidth}
\centering
  \includegraphics[width=1.\linewidth]{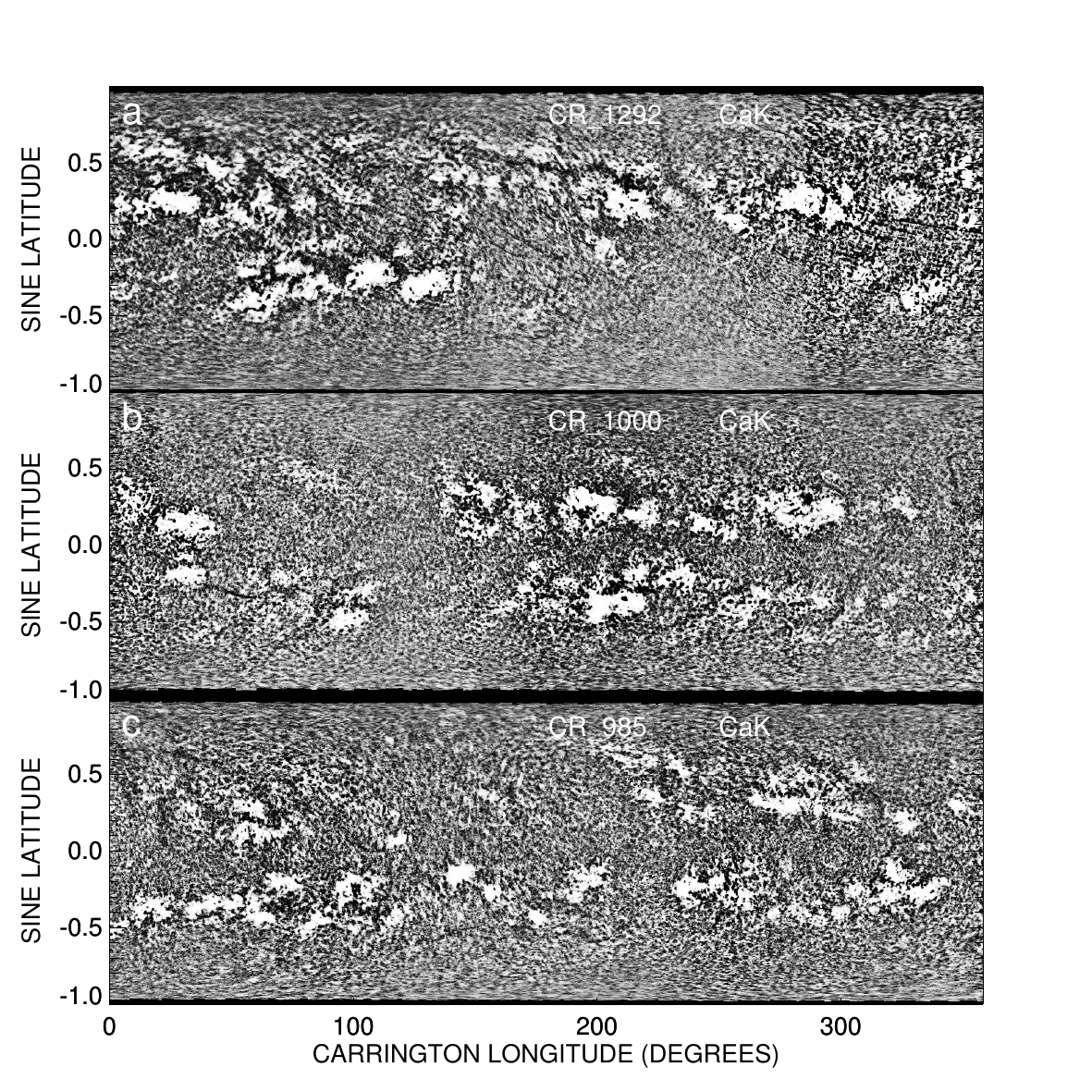}
 % \vspace{-.03\textwidth}
    \caption{Carrington maps generated from Kodaikanal observatory Ca~{\sc ii}~K full disc spectroheliograms.
   a) Carrington map starting at 8th April, 1950; b) Carrington map starting at 17th June, 1928; c) Carrington map starting at 5th May, 1927.}
  \label{fig:cr_CaK_new}
\end{figure}

\begin{figure}
%\vspace{-.2\textwidth}
\centering
  \includegraphics[width=1.\linewidth]{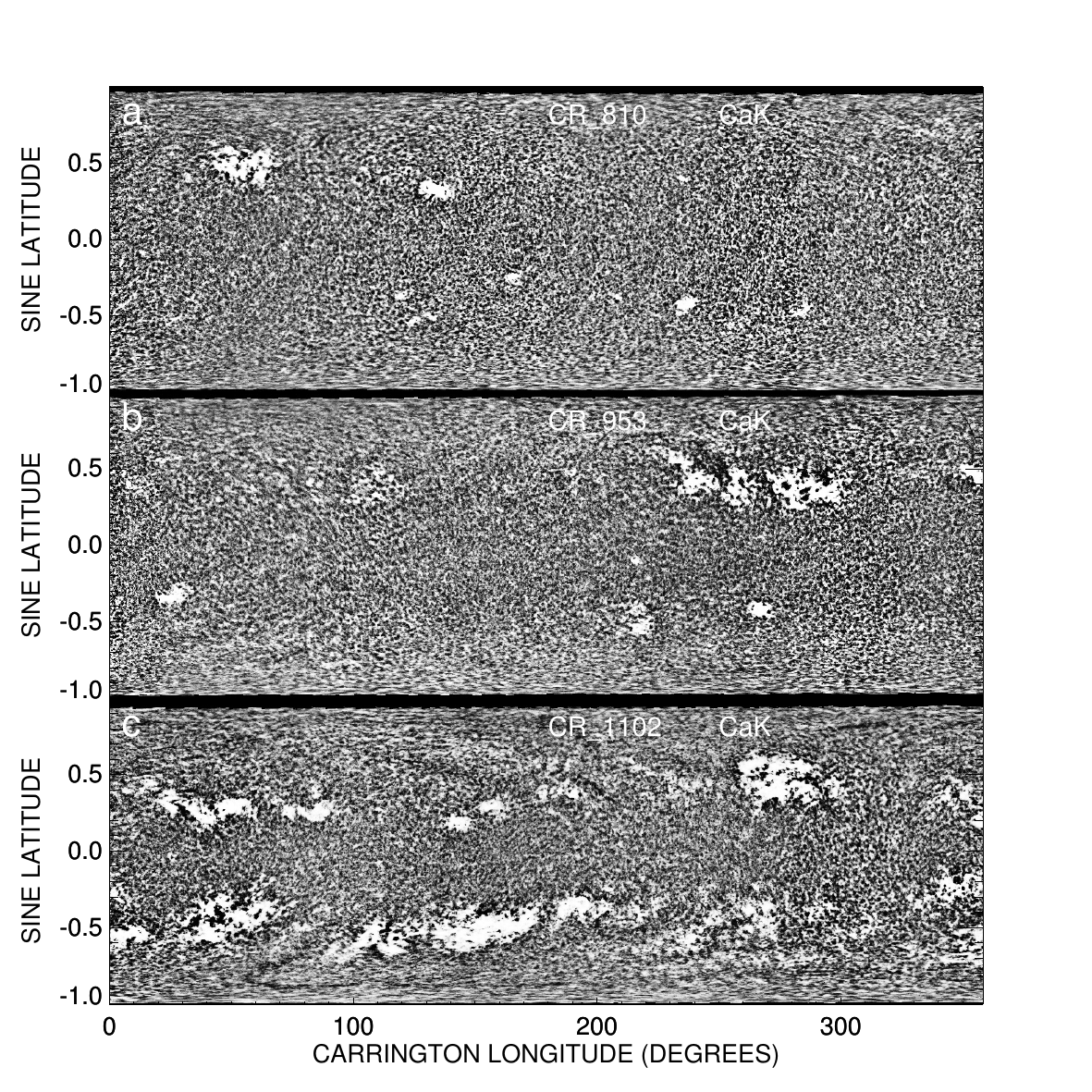}
 % \vspace{-.03\textwidth}
    \caption{Carrington maps generated from Kodaikanal observatory Ca~{\sc ii}~K full disc spectroheliograms.
   a) Carrington map starting at 10th April, 1914 which is at the minimum phase between cycle 14 and 15; b) Carrington map starting at 13th December, 1924 which is at the minimum phase between cycle 15 and 16; c) Carrington map starting at 30th January, 1936 which is at the beginning of rising phase of cycle 17. For all these three maps plages are mostly at higher latitudes $\approx 30^o$.}
  \label{fig:cr_CaK_new1}
\end{figure}

It has been pointed out earlier that the enhanced emission in Ca~{\sc ii}~K relates well with the magnetic field concentrations. Thus the Carrington maps do indicate the magnetic flux distribution over a solar rotation. In this article we presented Synoptic or Carrington maps (Figures~\ref{fig:cr_CaK}a-c) incorporating plage evolutions for different solar rotations over the past 100 years.  These maps will be particularly useful for the period when line of sight magnetograms are not available. Similar maps were generated by \citet{2011ApJ...730...51S} from MWO Ca~{\sc ii}~K data. First we will compare few representative cases of newly generated Carrington map from KSO data and that of MWO. Figures~\ref{fig:cr_CaK}~a,b,c can be compared with upper panel of Figure 4,  lower panel of Figure~7 and  lower panel of Figure 10 of \citet{2011ApJ...730...51S} respectively. The close resemblance of these maps provides a good validation of calibration, map generation and portray the quality of Kodaikanal data. The longitude band selected for each full disc image to generate the map was result of a trade off between feature sharpness and data gap  (Figure~\ref{fig:cr_CaK}b). However, apart from geometrical similarity a difference in contrast and visibility of network structures can be observed between the maps from KSO and MWO. This might have occurred because of the difference in digitisation and preprocessing of full disc images.  In Figures~\ref{fig:cr_CaK_new} a, b, c and Figures~\ref{fig:cr_CaK_new1} a, b, c we show few more examples of CR maps corresponding to different phases of a solar cycle as generated from Kodaikanal spectroheliograms without any data gaps. Enhanced Ca~{\sc ii}~K Carrington maps at later times (after 1990) show bright tilted periodic artefacts because of higher occurrence of scratches in the spectroheliograms. 

\begin{figure}
\centering
\vspace{-.11\textwidth}
  \includegraphics[width=1.\linewidth]{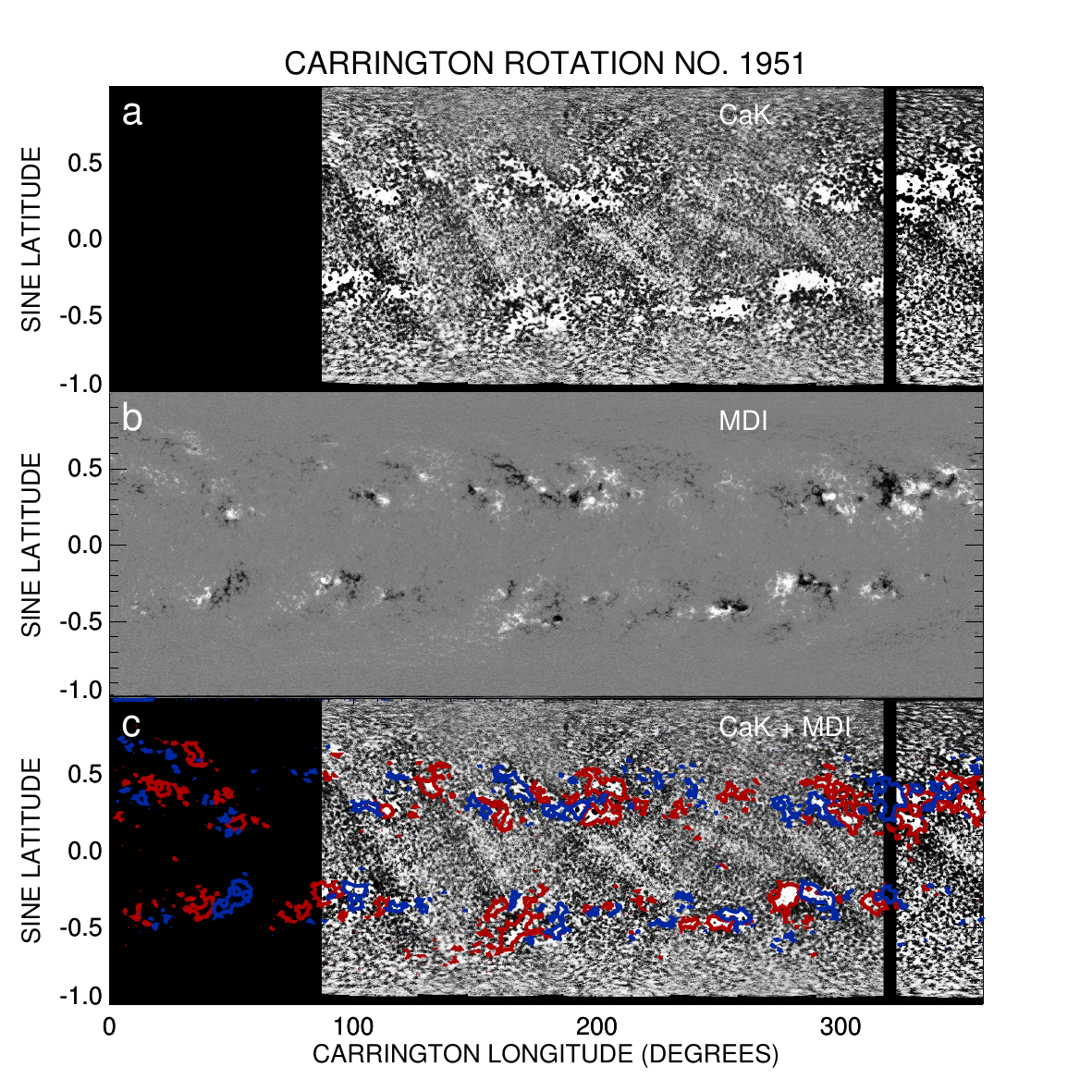}
  \caption{Overlap of plage structures and magnetic patches for Carrington rotation 1951.  a) Carrington map starting at 24th June, 1999 generated from Kodaikanal Observatory Ca~{\sc ii}~K full disc spectroheliograms; b) Carrington map generated from MDI full disc LOS magnetograms for the same rotation as that for Ca~{\sc ii}~K; c) Contours of large magnetic patches overplotted on (a). Red and blue depict towards and opposite to LOS direction. Positional correlation $\approx 86.3\%$.}
  \label{fig:cr_1951}
\end{figure}

\begin{figure}
\centering
\vspace{-.11\textwidth}
  \includegraphics[width=1.\linewidth]{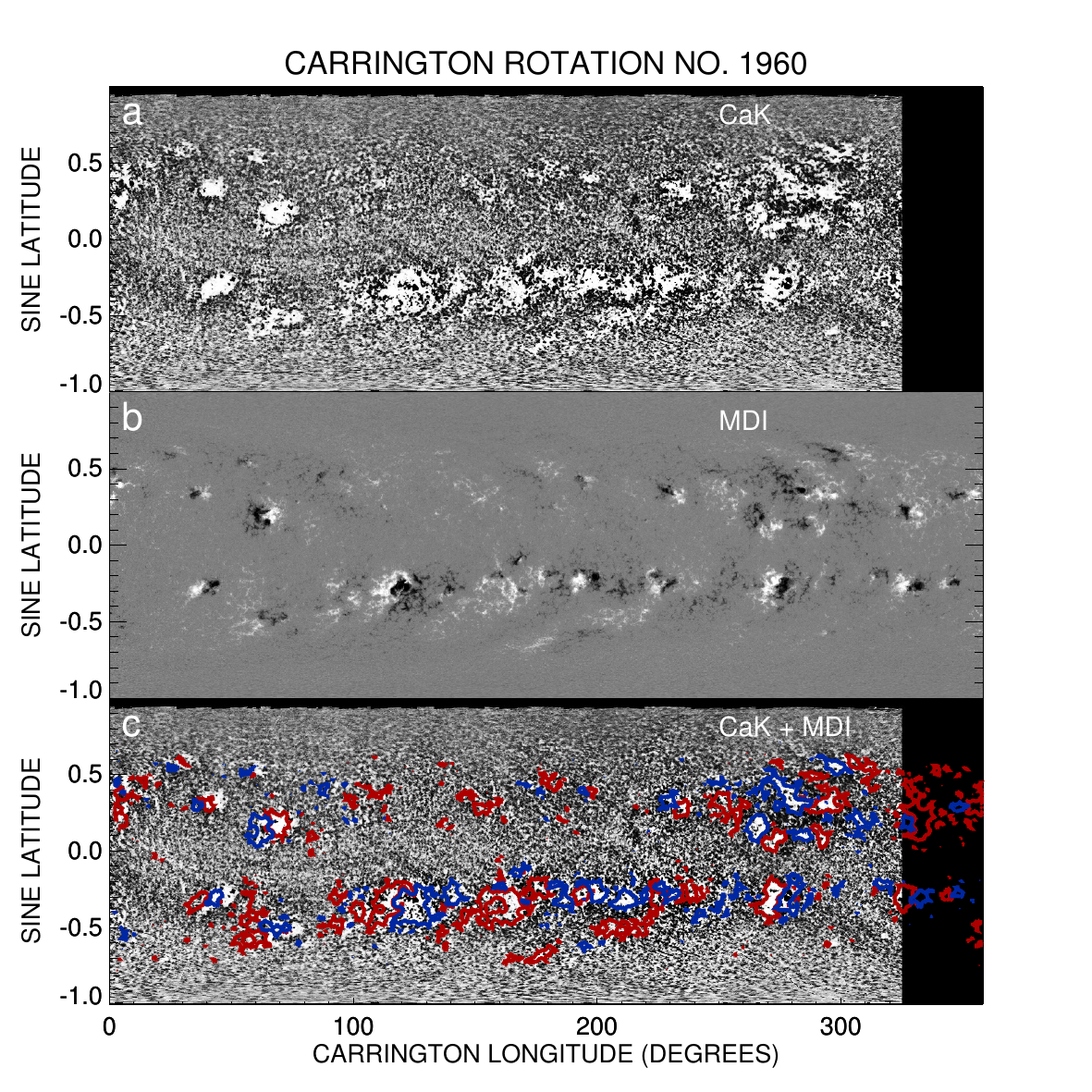}
  \caption{Overlap of plage structures and magnetic patches for Carrington rotation 1960. 
a) Carrington map starting at 25th February, 2000 generated from Kodaikanal Observatory Ca~{\sc ii}~K full disc spectroheliograms; b) Carrington map generated from MDI full disc LOS magnetograms for the same rotation as that for Ca~{\sc ii}~K; c) Contours of large magnetic patches overplotted on (a). Red and blue depict towards and opposite to LOS direction. Positional correlation $\approx 85.5\%$.}
  \label{fig:cr_1960}
\end{figure}

\begin{figure}
\centering
\vspace{-.11\textwidth}
  \includegraphics[width=1.\linewidth]{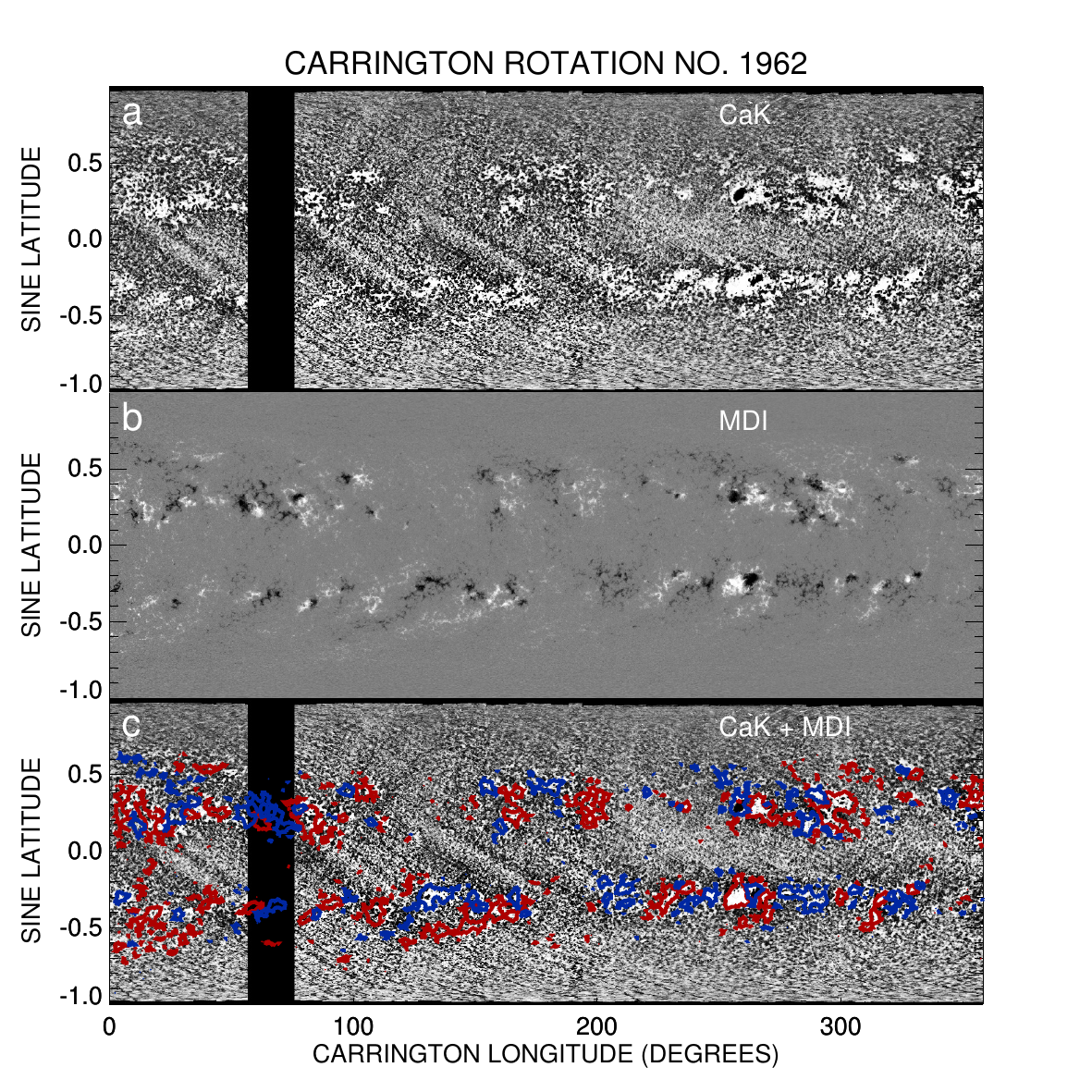}
  \caption{Overlap of plage structures and magnetic patches for Carrington rotation 1962. 
a) Carrington map starting at 19th April, 2000 generated from Kodaikanal Observatory Ca~{\sc ii}~K full disc spectroheliograms; b) Carrington map generated from MDI full disc LOS magnetograms for the same rotation as that for Ca~{\sc ii}~K; c) Contours of large magnetic patches overplotted on (a). Red and blue depict towards and opposite to LOS direction. Positional correlation $\approx 75.1\%$.}
  \label{fig:cr_1962}
\end{figure}

\begin{figure}
\centering
\vspace{-.11\textwidth}
  \includegraphics[width=1.\linewidth]{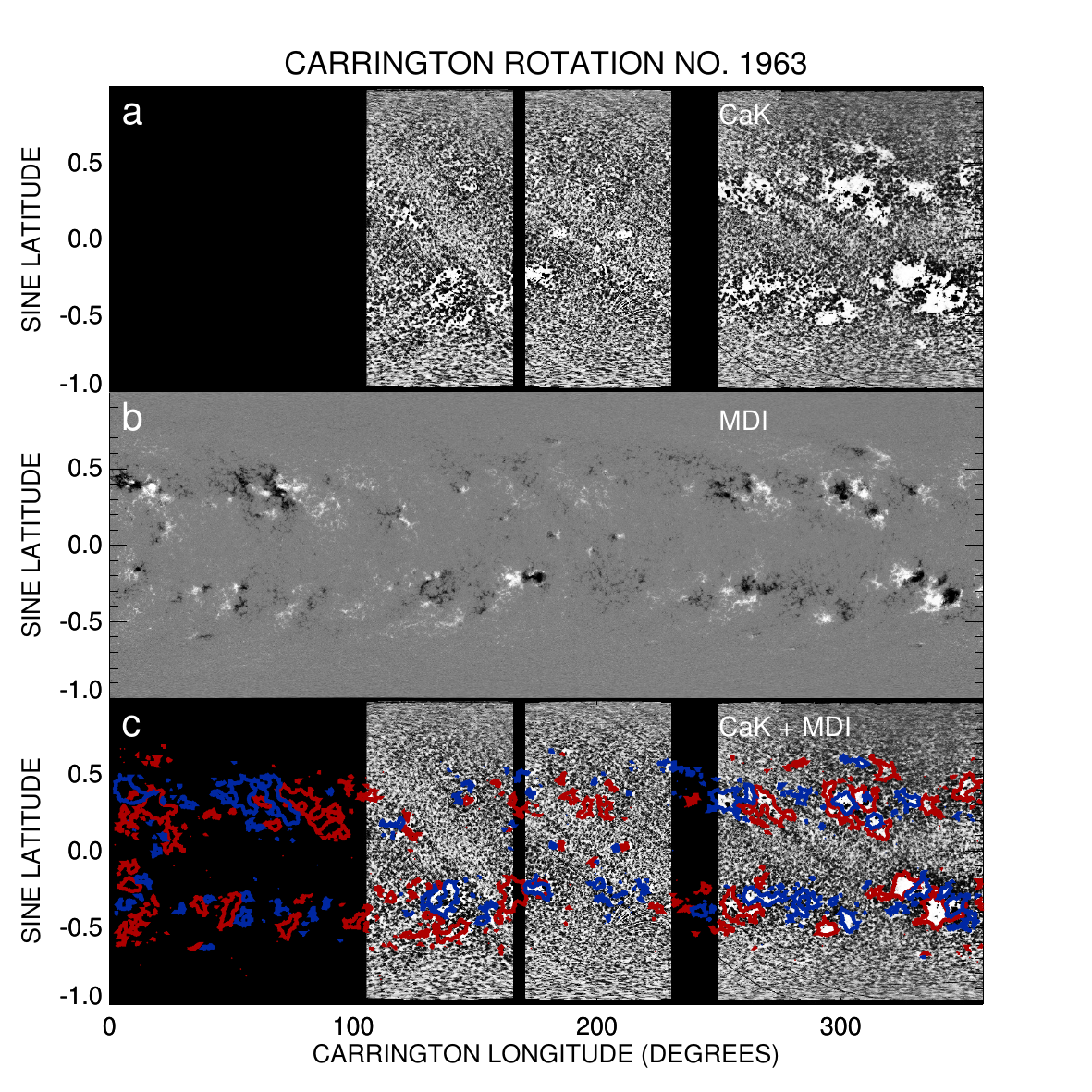}
  \caption{Overlap of plage structures and magnetic patches for Carrington rotation 1963.   a) Carrington map starting at 17th May, 2000 generated from Kodaikanal Observatory Ca~{\sc ii}~K full disc spectroheliograms; b) Carrington map generated from MDI full disc LOS magnetograms for the same rotation as that for Ca~{\sc ii}~K; c) Contours of large magnetic patches overplotted on (a). Red and blue depict towards and opposite to LOS direction. Positional correlation $\approx 92.3\%$.}
  \label{fig:cr_1963}
\end{figure}

Next we compare our Ca~{\sc ii}~K Carrington maps with Carrington maps of photospheric magnetic fields generated from full disc LOS magnetograms of  MDI on SoHO. Cross-correlation of all sample Carrington maps from Ca~{\sc ii}~K (KSO) and MDI showed maximum overlap at a relative x-shift and y-shift of about 5 pixels and 9 pixels respectively where size of the maps was $1571\times500$. This study reveals good  spatial correlation ($\geq75\%$) between magnetic patches and plage structures (Figures~\ref{fig:cr_1951},\ref{fig:cr_1960},\ref{fig:cr_1962},\ref{fig:cr_1963}). Carrington maps from 1907 till mid 2007 have been provided as supplementary material (also available at \url{ftp://ftp.iiap.res.in/subhamoy/cak_carrington_maps_kodaikanal/}). We also provide a time lapsed movie which includes all these maps for the period of 100 years in the link \url{ftp://ftp.iiap.res.in/subhamoy/cak_carrington_maps_kodaikanal/cak_video_1907_2007.gif} (available online). 
 

  \vspace{-.01\textwidth}
  \section{Conclusion}
We have used  Ca~{\sc ii}~K line images, recorded in the KSO, digitised and calibrated by our group, to construct Carrington maps and butterfly diagrams corresponding to the period of 1907-2007. The KSO data allows us to study the cyclic activity of the solar magnetic fields for longer duration ($>$100 years) in contrast to that from MWO which is for about 85 years \citep{2010SoPh..264...31B}. \citet{2011ApJ...730...51S} studied Carrington maps from 1915 till 1985 from the MWO archive, whereas the current study generated Carrington maps from 1907 to mid 2007 using images which were recorded at the KSO. These maps provide ways of tracking magnetic field distributions all the way back to 1907, a period from which direct magnetic field synoptic observations are not available. In the context of long term evolution of features on Sun at specific wavelength present study segmented plage structures, generated area cycle  and butterfly diagram from 100 years Ca~{\sc ii}~K KSO data. Derived century-long Ca~{\sc ii}~K synoptic maps further elaborated correlation with the same generated from space based magnetograms for an overlapping study period and confirmed their physical connection \citep{stix2004sun, 2011ApJ...730...51S}.
	 
	Figure~\ref{fig:butterfly_3d}  illustrated cyclic behaviour of the projected area for individual plages at different latitude bands.  It also gives hint about the relative delay in attaining cycle peak for individual plage area between different latitudes. So, this area scaled butterfly diagram gives some additional information for a greater time span (1907-2007) in contrast to the super synoptic map presented in \citet{2011ApJ...730...51S} with study period between 1915-1985.  Location overlap between plage structures and magnetic patches in turn validated the theoretical expectation of higher field strength correlated with less Ca~{\sc ii}~K absorption \citep{2016A&A...585A..40P}. Therefore the century-long Ca~{\sc ii}~K data can act as a proxy for understanding the locational evolution of magnetic patches. 
	
This study essentially demonstrated the importance of century-long Ca~{\sc ii}~K spectroheliograms to delineate evolutionary features from different perspectives and their possible correspondence. 
Observations about latitude dependent plage area variation can add some more insight to the conventional area cycle, being an aggregate. Along with the study of long term evolution some smaller time scale phenomena can also be analysed from this huge data. Rotation rate variation for different latitudes can be analysed from plage areas over 100 years \citep{kiepenheuer2012structure,singh}. Machine learning techniques can be applied on and validated through this huge data giving confidence about the prediction for future cycle characteristics \citep{SWE:SWE282}. We propose to address some of these aspects in our future studies.
%Major events such as flares in the past can be traced from this huge historical data from KSO and also recorded geomagnetic events can be correlated.  Theoretical speculations about the coupling of local and global dynamo models can eventually find some inputs from such data \citep{2014ApJ...793L...4P}.

 We hope that the result presented here and Carrington maps published online (\url{ftp://ftp.iiap.res.in/subhamoy/cak_carrington_maps_kodaikanal/}) will also help the global community to study different features at a greater depth and the new database available at \url{https://kso.iiap.res.in/data} will act as a valuable resource. Recently \citet{2016A&A...585A..40P} have attempted to create a homogeneous, long term series of pseudo-magnetograms using a combination of Ca~{\sc ii}~K line images and sunspot polarity measurements. We hope to pursue the same in near future from our KSO data. We further propose to combine the MWO and KSO data sets to produce a single improved set. This will also enable to fill up the data gaps in the two independent datasets. This cross-calibration and comparison will yield a data series with uniform quality. We would like to mention that further refinements, improvements in the calibration process and cross-calibration between different data series will be released time to time through our web portal at \url{https://kso.iiap.res.in/data} with marked versions and data products. %include raw image and Plage area N-S cycle
 
 \acknowledgements
 We would like to thank all the observers at Kodaikanal over 100 years for their contribution to build this enormous resource. The current high resolution digitisation process was initiated by Prof. Jagdev Singh and we thank him for his substantial  contribution to the project. We would like to thank many who have helped in the digitisation and calibration : Muthu Priyal, Amareswari, T. G. Priya, Ayesha Banu, A. Nazia, S. Kamesh, P. Manikantan, Janani, Manjunath Hegde, Trupti Patil, Sudha and staff members at Kodaikanal who also helped us in setting up the digitiser unit and the digitisation at Kodaikanal. We also thank the anonymous referee, whose valuable comments helped us to improve the presentation. 

\bibliographystyle{apj}

 \end{document}